\documentclass[]{raa}                  
\usepackage{graphicx,times}             
\input{epsf.sty}                        
\input{psfig.sty}                       

\begin{document}

   \title{Diagnose Physical Conditions Near the Flare Energy-release Sites from Observations of Solar Microwave Type III Bursts}

   \volnopage{Vol.00 (2015) No.0, 000--000}      
   \setcounter{page}{1}          

   \author{Baolin Tan \inst{1} \and Marian Karlick\'y \inst{2} \and Hana M\'esz\'arosov\'a \inst{2} \and Guangli Huang \inst{3}}
   \institute{Key Laboratory of Solar Activity, National Astronomical
Observatories of Chinese Academy of Sciences, Beijing 100012,
China, {\it bltan@nao.cas.cn}\\
   \and Astronomical Institute of the Academy of Sciences of the Czech Republic, CZ--25165 Ond\v{r}ejov, Czech Republic\\
        \and Purple Mountain Observatory of Chinese Academy of Sciences, Nanjing 210008, China\\ }

   \date{Received~~2009 month day; accepted~~2009~~month day}

\abstract{ In the physics of solar flares, it is crucial to
diagnose the physical conditions near the flare energy-release
sites. However, so far it is unclear how do diagnose these
physical conditions. Solar microwave type III burst is believed to
be a sensitive signature of the primary energy release and
electron accelerations in solar flares. This work takes into
account the effect of magnetic field on the plasma density and
developed s set of formulas which can be used to estimate the
plasma density, temperature, magnetic field near the magnetic
reconnection site and particle acceleration region, and the
velocity and energy of electron beams. We applied these formulas
to three groups of microwave type III pairs in a X-class flare,
and obtained some reasonable and interesting results. This method
can be applied to other microwave type III bursts to diagnose the
physical conditions of source regions, and provide some basic
information to understand the intrinsic nature and fundamental
processes occurring near the flare energy-release sites.
 }

   \authorrunning{Baolin Tan, Marian Karlick\'y, Hana M\'esz\'arosov\'a, Guangli Huang }            
   \titlerunning{Diagnostics of Microwave Type III Bursts}  

   \maketitle
\section{Introduction}
\label{sect:intro}

In solar flares, the energetic particles play a prominent role in
energy release, transmission, conversion, and emission
propagation. Estimation shows that energetic particles may carry
about 10 - 50\% of the total energy released in a typical X-class
flare (Lin \& Hudson, 1976). However, so far, there are many
unresolved problems on the solar energetic particles, especially
the energetic electrons, such as where is the site of electron
acceleration? What is the exact mechanism of particle
acceleration? How fast of the energetic electron beams flying away
from its source region? It is necessary to measure clearly the
physical conditions around the source region for answering the
above questions. These conditions include plasma density,
temperature, magnetic field, and energy of particle beams
(Zharkova et al. 2008).

Solar radio type III burst is believed to be a sensitive signature
of the energetic electron beams generated and propagated in the
corona. It is interpreted as caused by energetic electron beam
streaming through the background plasma at a speed of about 0.1 -
0.9$\mathrm{c}$ ($\mathrm{c}$ is the light speed) propagating away
from the acceleration site (e.g., Lin \& Hudson 1971, Lin et al.
1981, and a recent review in Reid \& Ratcliffe 2014). In meter or
longer wavelengths, solar radio type III bursts are regarded to be
generated by energetic electron beams propagating upward in open
magnetic field configurations, their frequency drift rates are
negative and be called normal type III bursts, their emission
source may locate above the flare energy-release site. But in
decimeter- or shorter wavelengths, the frequency drift rates of
type III bursts are always positive, we called them as microwave
type III bursts which regarded to be produced by the electron
beams propagating downward, their emission source may locate below
the flare energy-release site. Occasionally there will be observed
microwave type III burst pairs, which are composed of normal type
III branches with negative frequency drift rates and
reverse-sloped (RS) type III branches with positive frequency
drift rates simultaneously. Microwave type III pairs are explained
as produced by bi-directional electron beams and their source
regions are very close to the flare energy-release site where the
magnetic reconnection and particle acceleration take place.
Therefore, microwave type III bursts, including the microwave type
III pairs, are the most important tool to diagnose the physical
conditions around the flare energy-release sites. It may provide
the most important information for understanding the flare
triggering mechanism and particle acceleration.

The short lifetimes and very high brightness temperatures of
microwave type III bursts indicate that the emission should be
coherent processes. The first coherent mechanism is electron
cyclotron maser emission (ECM, Melrose \& Dulk 1982, Tang et al.
2012) which requires a strong background magnetic field:
$f_{ce}>f_{pe}$. $f_{ce}$ is electron gyro-frequency and $f_{pe}$
is plasma frequency. As for the microwave type III bursts, the
required magnetic fields should be $B>1000$\,G, too strong to
occur in solar corona. The second coherent mechanism is plasma
emission (PE, Zheleznyakov \& Zlotnik 1975) which is generated
from the coupling of two excited plasma waves at frequency of
2$f_{pe}$ (second harmonic PE), or the coupling of an excited
plasma wave and a low-frequency electrostatic wave at frequency of
about $f_{pe}$ (fundamental PE). As PE has no strong magnetic
field constraints, it is much easy to be the favorite mechanism of
type III bursts.

In PE mechanism, the emission frequency can be expressed
\begin{equation}
f\approx 9sn_{e}^{\frac{1}{2}}.
\end{equation}
The units of $f$ and $n_{e}$ are Hz and m s$^{-3}$, respectively.
$s$ is the harmonic number, $s=1$ is the fundamental PE while
$s=2$ is the second harmonic PE. Generally, the fundamental PE has
strong circular polarization, while the second harmonic PE always
has relatively weak circular polarization.

The frequency drift rate is the most prominent observed parameter
of microwave type III bursts, which is connected to many physical
conditions in the source region. From Equation (1), we derived:
\begin{equation}
D=\frac{df}{dt}=\frac{f}{2}(\frac{\partial n_{e}}{n_{e}\partial
t}+\frac{\partial n_{e}}{n_{e}\partial l}\cdot \frac{\partial
l}{\partial t})=\frac{f}{2}(\frac{1}{t_{n}}+\frac{v_{b}}{H_{n}})
\end{equation}
The relative frequency drift rate is expressed,
\begin{equation}
\bar{D}=\frac{df}{fdt}=\frac{1}{2t_{n}}+\frac{v_{b}}{2H_{n}}
\end{equation}
$H_{n}=\frac{n_{e}}{\partial n_{e}/\partial l}$ is the plasma
density scale length along the electron beams. $H_{n}>0$ or
$H_{n}<0$ means the density increasing or decreasing along the
electron beam. $t_{n}=\frac{n_{e}}{\partial n_{e}/\partial t}$ is
the plasma density scale time. $t_{n}>0$ or $t_{n}<0$ means the
density increasing or decreasing with respect to time.
$\frac{\partial l}{\partial t}=v_{b}$ is the velocity of electron
beams.

Equation (2) and (3) shows that frequency drift rate is composed
of two parts. The first part is related to the temporal variation
of plasma density, and the second part is related to the spatial
variation of plasma density which is caused by plasma density
gradient and the motion of electron beams.

In flare impulsive phase, fast magnetic reconnection may produce
plasmas inflow into the reconnecting region and result in plasma
density increasing, $t_{n}>0$. In postflare phase, magnetic
reconnection becomes very weak, and there is no obviously
variation of the plasma density, $t_{n}\rightarrow \infty$ and the
first term of Equation(2) and (3) can be neglected.

$H_{n}$ depends on plasma density distributions (Dulk 1985). So
far, the most common plasma density models of the quiet solar
atmosphere are Baumbach-Allen model (Allen 1947) and Newkirk model
(Newkirk, 1967). When applying these models to the active region,
it is always necessary to multiply by a number from 3 to 50.
Naturally, this method has a big uncertainty. By using the
broadband radio spectral observations of the solar eclipse, Tan et
al. (2009) obtained an improved semi-empirical model of the
coronal plasma density. However, the above models are expressed as
functions of height from the solar surface. It is difficult to
apply them to a microwave type III burst at a certain frequency
when the height of the source region is not known. Under the
assumption of solar static and barometric atmosphere, $H_{n}$ can
be estimated as $H_{n}=-\frac{k_{B}T}{m_{0}g}$ (Benz et al. 1983).
$k_{B}$ is the Boltzmann constant, $g$ the solar gravitational
acceleration ($\sim$274 ms$^{-2}$ near the solar surface), and
$m_{0}$ the average mass of ions in the solar chromosphere and
corona. The minus sign indicates the density decreasing with
respect to the height. The barometric model is only valid in the
quiet corona and not appropriate for the type III-emitting source
region that cannot be in hydrodynamic equilibrium because of the
substantial heating and plasma flows. The generated microwaves are
subject to free-free absorption when they propagates in the
ambient plasma, they will be strongly absorbed. Assuming the
plasma is isothermal and barometric, then the density scale length
corresponding to the optical depth unity is about,
$H_{n}\approx6.67\times10^{14}\frac{T^{3/2}}{f^{2}}$ (m) for the
fundamental emission, and
$H_{n}\approx1.07\times10^{16}\frac{T^{3/2}}{f^{2}}$ (m) for the
harmonic emission (Dulk 1985, Stahli \& Benz 1987). But when
applying this expression to estimate the electron beam velocity,
we find $v_{b}>1.0$ c in some events, and $v_{b}<0.1$ c in other
events. These results are unreasonable. Aschwanden \& Benz (1995)
proposed a combined model by assuming a power-law function in the
lower corona and an exponential function in the upper corona.
However, as the above models have not considered the effect of
magnetic fields, they can not reflect the relationship between the
physical conditions and the properties of the microwave type III
bursts.

This work plans to propose a new method to diagnose the physical
conditions around the flaring energy-release sites from
observations of solar microwave type III bursts. Section 2
introduces the main observable parameters of microwave type III
bursts. In Section 3, we derived a new expression of the plasma
density scale length ($H_{n}$) from the full MHD equation which
contains magnetic pressure force. On this basis, we proposed a set
of formulas to estimate the plasma density, temperature, magnetic
field around the primary source region of solar flares, and the
velocity and energy of the energetic electron beams. In Section 4,
we apply the above method to diagnose the source region of three
groups of microwave type III bursts pair in the postflare phase of
a X-class flare. Finally, conclusions are summarized in Section 5.

\section{Observing Parameters of Microwave Type III Bursts}
\label{sect:Obs}

Recently, there are many solar radio spectrometers having been
operated at frequency of microwave range with high frequency and
time resolutions. For example, the Chinese Solar Broadband Radio
Spectrometers at Huairou (SBRS) operates with dual circular
polarization (left- and right-handed circular polarization) in the
frequency range of 1.10 -- 2.06 GHz, 2.60 -- 3.80 GHz, and 5.20 --
7.60 GHz, their cadence is 5 -- 8 ms, and the frequency resolution
is 4 -- 20 MHz (Fu et al. 1995, 2004, Yan et al. 2002). The
Ond\v{r}ejov radiospectrograph in the Czech Republic (ORSC)
operates at frequencies of 0.80-5.00\,GHz, its cadence is 10 ms
and the frequency resolution is 5-12 MHz (Ji\v{r}i\v{c}ka et al.
1993).

From the observations of the microwave type III bursts, we can
derived several observable parameters:

(1) Start frequency ($f_{st}$), defined as the frequency at the
start point of the microwave type III burst with emission
intensity exceeding the background significantly.

(2) Burst lifetime ($\tau$), defined as the time difference
between the start and end of an individual type III burst.

(3) Central frequency ($f_{0}$), defined as the frequency at the
midpoint of type III burst.

(4) Frequency drift rates ($D=\frac{df}{dt}$), defined as slopes
of type III burst on the spectrograms. $f$ is the frequency at the
ridge crest along the type III burst at different time. Generally,
frequency drift rate can be derived by using cross-correlation
analysis at each two adjacent frequency channels.

(5) The relative frequency drift rates ($\bar{D}$), defined as
$\bar{D}=\frac{df}{fdt}\approx\frac{D}{f_{0}}$.

\section{Diagnostics to the Flare Energy-release Site}

In order to understand the process of microwave type III bursts,
we construct a physical scenario to demonstrate roughly the
possible processes of magnetic reconnection, particle
accelerations and propagations associated to microwave type III
bursts in solar flares (Fig. 1). Here magnetic reconnection may
take place in a cusp configuration above and very close to the top
of flaring loop similar as the Masuda-like flare (Masuda et al.
1994, etc.), or in the current sheet above and beyond flare loops
triggered by the tearing-mode instability (Kliem et al. 2000,
etc). In some flares the reconnections are more dominated in cusp
configuration and in other flares in current sheets. Electrons are
accelerated in the reconnection site (source region), propagate
upward or downward, and produce normal or RS type III bursts.

\begin{figure}[ht]
\begin{center}
   \includegraphics[width=9.2 cm]{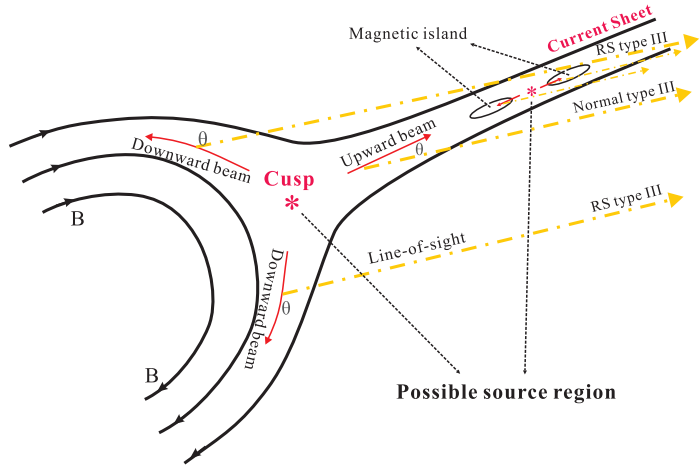}
\caption{Schematic diagram of the formation of radio type III
burst pairs. Red arrows indicate the path of the energetic
electron beams, and the yellow arrows indicate the emission
direction.}
\end{center}
\end{figure}

In this section, a new diagnostic method can be derived in
framework of the above physical scenarios.

\subsection{Constraints of the Emission Process}

As we mentioned in Section 1, the most possible emission mechanism
of microwave type III bursts is the plasma emission. There are
some constraints to this mechanism for a certain microwave type
III bursts (Cairns \& Melrose 1985, Robinson \& Benz 2000):

(1) Magnetic field. $f_{ce}<f_{pe}$ and $f>f_{pe}$ requires a
relatively weak magnetic field $B<3.57\times10^{-7}f$ (G). Here,
the unit of $f$ is Hz. For example, the magnetic field must be
$B<1070$\,G at frequency of 3\,GHz. For most cases, this condition
is easy to agree.

(2) Emission Frequency. PE mechanism is a three-wave interaction
process, which must satisfy frequency condition
$f_{T}=f_{L}+f_{3}$. Here $T$, $L$, and $3$ represent the
transverse radiation, Langmuir wave, and the third wave,
respectively. When the third wave is an ion-sound wave it will
produce fundamental emission; and a second harmonic emission will
produce when the third wave is a secondary Langmuir wave.

(3) Beam direction. The three-wave interaction requires
wave-vector conservation $k_{T}=k_{L}+k_{3}$. Here, $k_{L}$ is
almost at the electron beams direction, while $k_{3}$ have broad
distributions. As for the fundamental emission, $k_{3}$ represents
an ion-sound wave and $k_{3}\ll k_{L}$, the emission ($k_{T}$) is
approximately at the same direction of electron beams. As for the
second harmonic emission, $k_{3}$ represents a secondary Langmuir
wave, $k_{T}$ may obviously deviate from the electron beams.

(4) Beam velocity. Not all electron beams can generate Langmuir
waves to produce a certain type III burst, actually only the beam
with positive derivative distribution function in velocity space
can produce Langmuir waves by bump-in-tail instability. The
positive derivative distribution function must be formed on the
total distribution function (background Maxwellian distribution +
beam distribution), and this means that the velocity of electron
beam must exceed a lower limit which is above the thermal velocity
of the background Maxwellian distribution, especially for the low
density beams:
\begin{equation}
v_{b}>3.2\times10^{2}(n_{e}\tau)^{1/3}=v_{L}
\end{equation}
The above equation is derived from the relationship $\tau<t_{d}$.
$\tau$ is the observed lifetime of microwave type III burst.
$t_{d}=3.1\times10^{-8}\frac{v_{b}^{3}}{n_{e}}$ is the collisional
deflected time of electron beams, here the units of $v_{b}$ and
$n_{e}$ are $m\cdot s^{-1}$ and $m^{-3}$, respectively (Benz et
al. 1992). The lower limit indicates that low energy electron beam
will be dissipated in dense plasma and cannot produce coherent
emission.

At the same time, the dispersion relation of PE emission also
requires an upper velocity limit:
\begin{equation}
v_{b}<\frac{3}{2}\sqrt{\frac{m_{i}}{\gamma m_{e}}}V_{0}=v_{H}
\end{equation}
Here $\gamma=1+3\frac{T_{i}}{T_{e}}\approx 2$, $V_{0}$ is the
electron thermal velocity, $v_{H}$ is a critical velocity
($\approx40-50V_{0}$), $m_{e}$ and $m_{i}$ are electron and mean
ion mass, $T_{i}$ and $T_{e}$ are ion and electron temperatures,
respectively. The upper velocity limit indicates that very high
energy electron beam will not produce plasma emission when
propagating in cold plasmas. In very hot plasma, for example, when
$T_{e}\approx3\times10^{6}$ K, $v_{H}\approx c$. At such case, the
upper limit is the light speed. The full velocity constraint for
PE-emitting electron beam is $v_{L}<v_{b}<v_{H}$. This constraint
is very meaningful for estimating the magnetic field near source
regions. We will demonstrate it in Section 3.5.

\subsection{Plasma Density and Temperature}

The emission of microwave type III burst generates close to the
flare primary source region. By using the observed start frequency
($f_{st}$), the plasma density near the start site of the electron
beam can be obtained from Equation (1):
\begin{equation}
n_{st}=f_{st}^{2}/81s^{2}, (m^{-3}).
\end{equation}

The temperature can be derived approximately from the ratio of
soft X-ray (SXR) intensities at two energy band observed by GOES
satellite after subtracting the background by using the Solar
Software (Thomas et al. 1985). Applying this method, we obtained
the temperature profiles in each flare. From these profiles, we
can get the temperature at the time of microwave type III burst.
However, it is important to note that the above temperature is
only related to the hot flaring loops which is possible different
from the real source region of microwave type III bursts.

\subsection{Frequency Drift Rates}

Equation (2) and (3) indicate that the frequency drift rate of
microwave type III burst is composed of two parts. When the
temporal change is neglected, the frequency drift rate only
depends on the ratio between the beam velocity ($v_{b}$) and the
density scale length ($H_{n}$) along the path of source (Benz et
al. 1983). Therefore, $H_{n}$ is a key factor to estimate the beam
velocity from frequency drift rate of microwave type III bursts.
However, as we mentioned in Section 1, there is a considerable
uncertainty to determine the value of $H_{n}$ by using the
existing methods. All the above methods have not considered the
effect of magnetic fields on the plasma density and its
distribution.

Actually, magnetic field is a key factor in flare energy-release
sites, which compresses and confines plasma, and dominates the
plasma density and its distribution. As a consequence, $H_{n}$ is
also dominated by the magnetic field. In this section, we try to
deduce an expression of $H_{n}$ from a full MHD equation including
plasma thermal pressure force ($\nabla p$), Lorentz force
($\vec{j}\times\vec{B}$), and gravitational force ($\rho\vec{g}$),

\begin{equation}
\frac{d(\rho \vec{v})}{dt}=-\nabla
p+\vec{j}\times\vec{B}+\rho\vec{g}
\end{equation}

In quasi-equilibrium, $\frac{d(\rho \vec{v})}{dt}\sim0$, $\nabla
p\sim\frac{n_{e}k_{B}T}{H_{n}}$. The Lorentz force can be
decomposed into:

\begin{equation}
\vec{j}\times\vec{B}=\frac{1}{\mu_{0}}(\vec{B}\cdot\nabla)\vec{B}-\nabla(\frac{B^{2}}{2\mu_{0}})
\end{equation}
Furthermore, the first and second terms in the right side can be
expressed respectively as,

\begin{equation}
\frac{1}{\mu_{0}}(\vec{B}\cdot\nabla)\vec{B}=\frac{d}{ds}(\frac{B^{2}}{2\mu_{0}})\vec{s}+\frac{B^{2}}{\mu_{0}R_{c}}\vec{s}
\end{equation}

\begin{equation}
-\nabla(\frac{B^{2}}{2\mu_{0}})=-\frac{d}{ds}(\frac{B^{2}}{2\mu_{0}})\vec{s}-\nabla_{\perp}(\frac{B^{2}}{2\mu_{0}})\vec{n}.
\end{equation}

Here, the unit vector $\vec{s}$ is along the magnetic field line,
and $\vec{n}$ is normal to the magnetic field lines, and $R_{c}$
is the curvature radius of the magnetic field lines.

Substituting Equation (9) and (10) into (8), then, we have
\begin{equation}
\vec{j}\times\vec{B}=\frac{B^{2}}{\mu_{0}R_{c}}\vec{s}-\nabla_{\perp}(\frac{B^{2}}{2\mu_{0}})\vec{n}.
\end{equation}
The first term of Equation (11) is the magnetic tension force
which is acting along the direction of magnetic field, has
resultant effect only when the magnetic field line is curved. The
second term is the magnetic pressure force which is acting at the
direction perpendicular to the magnetic field lines, and this
force will compress and confine the plasmas (Priest 2014). Any
changes of the magnetic field strength will lead to a
corresponding variation in plasma density. Therefore plasma
density is dominated by the magnetic field. In the actual
condition near the flare energy-release site, there also has
magnetic gradient between different points along the magnetic
field lines, and this will lead to a density gradient between the
different points along the magnetic field. Substituting Equation
(11) into Equation (7), we have,

\begin{equation}
\frac{d(\rho
\vec{v})}{dt}=-[\nabla_{\parallel}P-\frac{B^{2}}{\mu_{0}R_{c}}+m_{0}n_{st}g\cos\theta]
\textbf{s}-[\nabla_{\perp}P+\nabla_{\perp}(\frac{B^{2}}{2\mu_{0}})-m_{0}n_{st}g\sin\theta]\textbf{n}
\end{equation}
Here, the plasma thermal pressure $P=n_{st}k_{B}T_{e}$. Fig. 1
indicates that the electron beams are mainly propagating along the
magnetic field lines, we may let the longitudinal component
approached zero,
$\nabla_{\parallel}(n_{st}k_{B}T_{e})\approx\frac{n_{st}k_{B}T_{e}}{H_{n}}$.
$H_{n}$ is the plasma density scale length along the magnetic
field lines,
\begin{equation}
H_{n}=\frac{k_{B}T}{\frac{B^{2}}{n_{st}\mu_{0}R_{c}}-m_{0}g\cos\theta}.
\end{equation}

In barometric atmosphere model, $\vec{j}\times\vec{B}\rightarrow
0$, we can derive the barometric density scale length
$H_{n}=-\frac{k_{B}T}{m_{0}g}$ directly from Equation (7).
Actually, when the magnetic field is very weak $B\rightarrow0$, or
in a homogeneous magnetic field $R_{c}\rightarrow\infty$, the
first term of the denominator vanishes, and assume
$\theta\rightarrow0$, Equation (13) will degenerate to be the
barometric format. For example, the radio type III burst at meter
or longer wavelength is generally explained as the propagation of
electron beam in the open magnetic field in upper corona where the
magnetic field is very weak and the curvature radius of the
magnetic field lines is very longer, the barometric model can
explain the main properties of radio type III bursts very well.

As for the microwave type III burst, its source region is very
close to the acceleration regions and very low. Here the magnetic
field is relatively strong, for example, with a possible typical
values of parameters, $B=$ 20 G, $n_{st}=10^{10}$ cm$^{-3}$, and
$R_{c}=10^{4}$ km, then $\frac{B^{2}}{n_{st}\mu_{0}R_{c}}\sim$ 100
$m_{0}g$, that means it is very easy to have the relation
$\frac{B^{2}}{n_{st}\mu_{0}R_{c}}\gg m_{0}g$. This fact implies
that magnetic field does play a dominant role to the plasma
density scale length near acceleration regions. Then, Equation
(13) can be approximated,

\begin{equation}
H_{n}\approx\frac{\mu_{0}n_{st}k_{B}T}{B^{2}}R_{c}=\frac{1}{2}\beta_{p}\cdot
R_{c}.
\end{equation}
$\beta_{p}=\frac{n_{st}k_{B}T}{B^{2}/(2\mu_{0})}$ is the plasma
beta defined as the ratio between the plasma thermal pressure and
the magnetic pressure. Equation (14) indicates that the plasma
density scale length ($H_{n}$) can be replaced by measuring the
configuration of the magnetic field ($R_{c}$).

\subsection{Electron Beams}

The electron beams associated to microwave type III bursts carry
the important information of source region, acceleration
mechanism, and the trigger mechanism of solar flares. Here, the
beam velocity ($v_{b}$) is a key factor to reflect the above
problems. From Equation (3) and (14), the beam velocity can be
derived,

\begin{equation}
v_{b}=\frac{\mu_{0}n_{st}k_{B}T}{B^{2}}(2\bar{D}-t_{n}^{-1})R_{c}
\end{equation}

The plasma density scale time ($t_{n}$) can be derived from the
drift rate of the start frequency ($df_{st}/dt$):
$t_{n}=(2\frac{df_{st}}{f_{st}dt})^{-1}$. Observations show that
the maximum drift rate of start frequency is only several MHz per
second, and the relative drift rate is
$\frac{df_{st}}{f_{st}dt}\sim 10^{-3}$ s$^{-1}$, $t_{n}^{-1}\sim
10^{-3}$ s$^{-1}$. The observation shows that the value of
$\bar{D}$ is ranging from 0.3 s$^{-1}$ to 5.3 s$^{-1}$. Therefore,
we have $\bar{D}\gg t_{n}^{-1}$, and the term of plasma density
time scale can be neglected. Equation (15) can be approximated,

\begin{equation}
v_{b}\approx\frac{2\mu_{0}n_{st}k_{B}T}{B^{2}}\bar{D}R_{c}
\end{equation}

Here, the magnetic field $B$ is not known. Section 3.5 will
discuss the method of estimating magnetic field. The electron beam
energy is $E_{b}\approx 256\frac{\beta^{2}}{\sqrt{1-\beta^{2}}}
(keV)$. Here, $\beta=v_{b}/c$.

\subsection{Estimation of the Magnetic Field}

As mentioned in Section 3.1, there are velocity constraints for
PE-emitting electron beams. With these constraints and Equation
(15) and (16), an estimation of magnetic field in source regions
can be derived. The GOES SXR observations show that the
temperature near the source regions exceeds $10^{7}$ K, the upper
velocity limit is: $v_{b}<c$ (Equation 5). Then the lower limit of
magnetic fields can be obtained,
\begin{equation}
B_{L}>3.402\times10^{-19}(n_{st}T|\bar{D}R_{c}|)^{1/2}.
\end{equation}

From the lower limit of beam velocity (Equation 4), an upper limit
of the magnetic field can be obtained,
\begin{equation}
B_{H}<3.293\times10^{-16}[\frac{n_{st}T|\bar{D}R_{c}|}{(n_{st}\tau)^{1/3}}]^{1/2}.
\end{equation}

The unit of magnetic field strength ($B$) is Tesla. Then the
magnetic field near the start site of the relates electron beams
should be in the range of $B_{L}<B<B_{H}$. The median of the upper
and lower limits should be the best estimator of the real magnetic
field strength $B\sim\frac{1}{2}(B_{L}+B_{H})$, the difference
between $B_{L}$ and $B_{H}$ can be regarded as the error limit.

The magnetic field curvature radius $R_{c}$ can be obtained from
imaging observations. By using Equation (17) and (18), we may
obtain the magnetic field $B$. Furthermore, the beam velocity
($v_{b}$) can be obtained from Equation (15) or (16). Together
with the results of Section 3.2, we obtain a full diagnostic
method of the flare energy-release site, include plasma density,
temperature, magnetic field, and velocity and energy of electron
beams.

\section{An Example}

We take the microwave type III burst pairs at frequency of 2.6 -
3.8 GHz observed by SBRS on 2006 December 13 as an example to test
the above method in diagnosing the physical conditions of primary
energy release. A type III burst pair composed of a group of
normal type III bursts (normal branch) with negative frequency
drift rates and a group of reversed slope type III bursts (RS
branch) with positive frequency drift rates (Tan et al. 2015).
Fig. 2 presents spectrograms of three group of microwave type III
pairs in a long-duration powerful X3.4 flare. Table 1 lists their
observing properties. Comparing to the general microwave type III
bursts, the microwave type III pair burst may provide much more
information of the flare energy-release sites. The separate
frequency ($f_{x}$) between the normal and RS branches can be used
to replace the start frequency ($f_{st}$) in Equation (6) to
obtain the plasma density ($n_{x}$). By applying Equation (16) and
(17) to the normal and RS branches, we can obtain the estimation
of magnetic fields near the acceleration region.

\begin{figure*}[ht] 
\begin{center}
   \includegraphics[width=7.3 cm, height=8.5 cm]{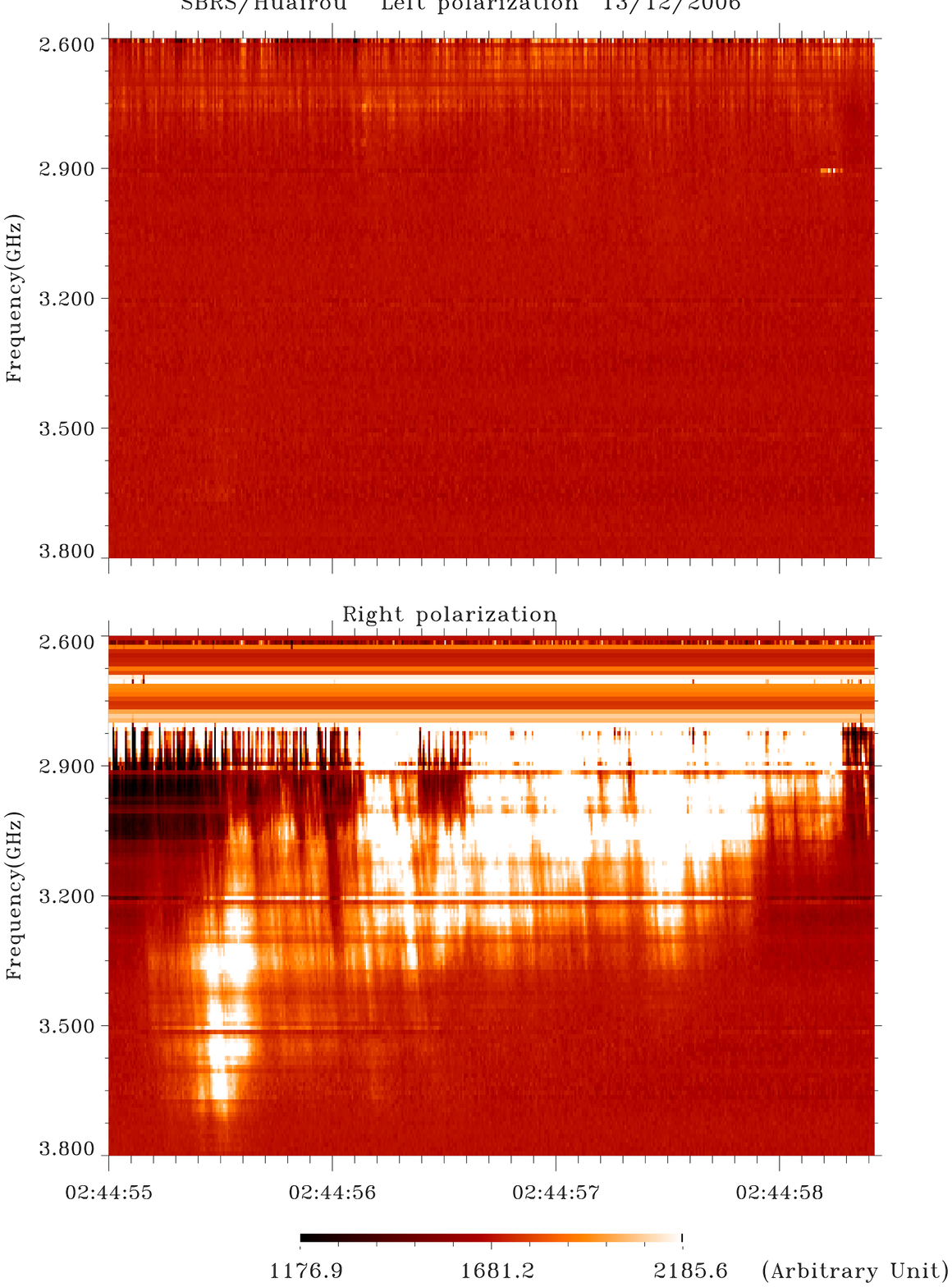}
   \includegraphics[width=7.3 cm, height=8.5 cm]{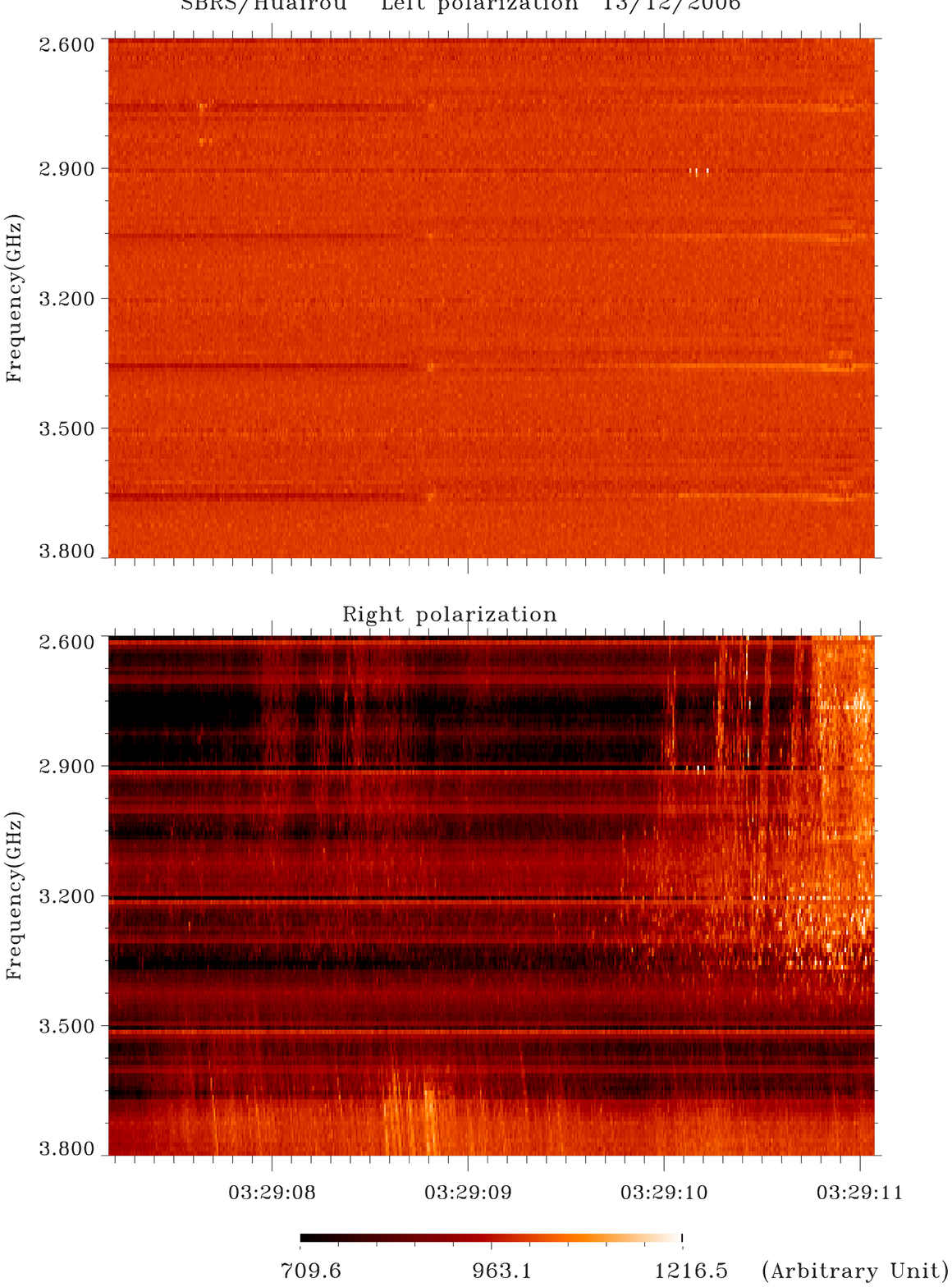}
   \includegraphics[width=7.3 cm, height=8.5 cm]{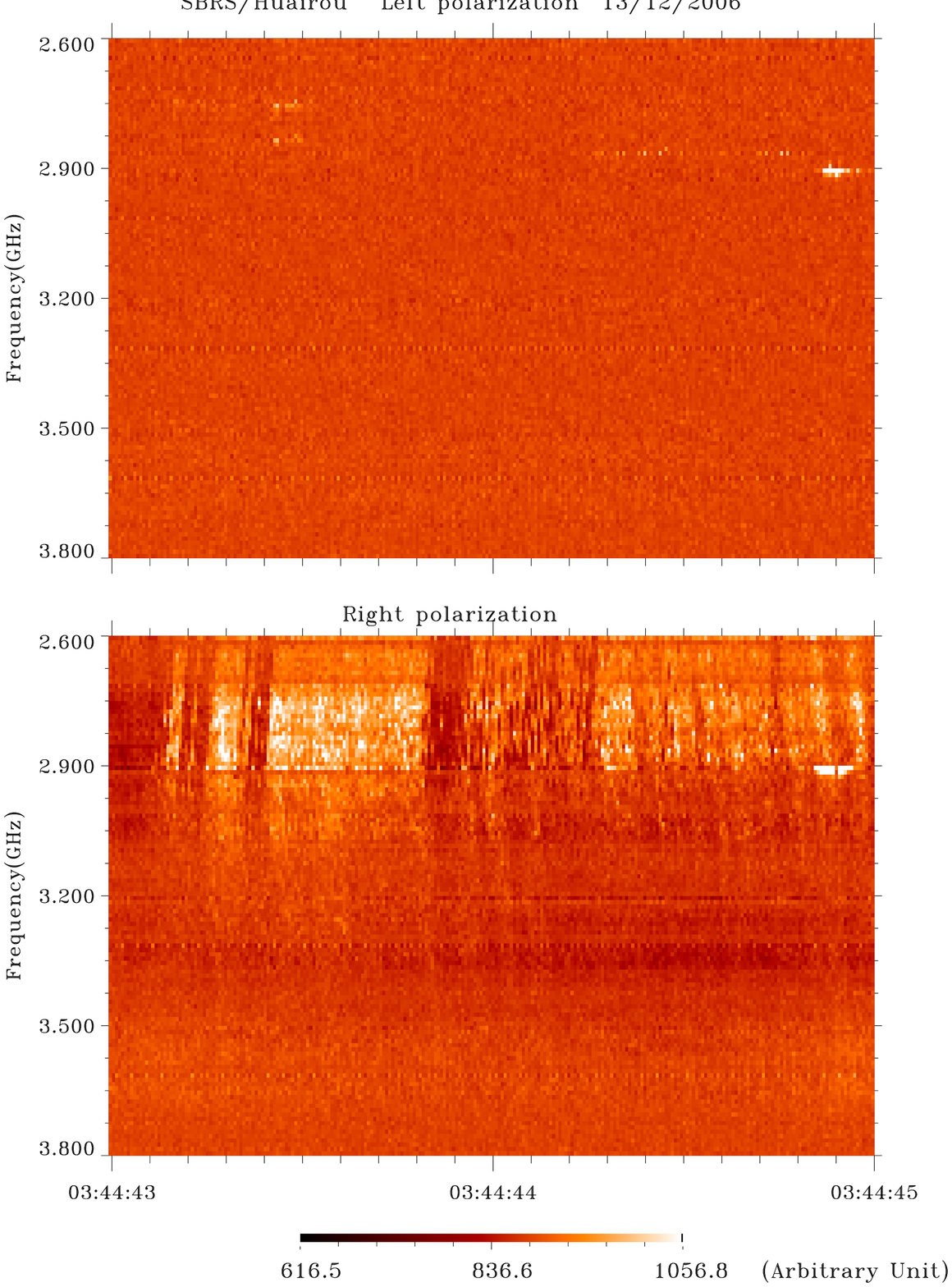}
\caption{Three groups of microwave type III pairs at 02:44:56\,UT
(A), 03:29:09\,UT (B), and 03:44:43\,UT (C) observed at
SBRS/Huairou on 2006 December 13. The right-bottom panel shows the
the whole flare process profiled by GOES soft X-ray at 1-8 \AA~
and microwave emission at 3.75 GHz.}
\end{center}
\end{figure*}

\begin{table}
\begin{center}
 \caption[]{Estimation of the physical parameters in the source region of type III pairs occurring in the postflare phase of the X3.4 flare on 2006 December 13.}
 \begin{tabular}{ccccccccccc}
  \hline\noalign{\smallskip}
               & Parameter                                        & Type III pair A   & Type III pair B  &  Type III pair C \\\hline\noalign{\smallskip}
  Observation  & Start time (UT)                                  &     02:45         &   03:29          &   03:45          \\
               & Lifetime: $\tau$ (s)                             &      0.16         &    0.10          &    0.07          \\
               & Separate frequency: $f_{x}$ (GHz)                &      2.92         &    3.42          &    2.90          \\
               & Frequency Gap: $\delta f$ (MHz)                  &      50           &    260           &    10            \\
               & Normal III: $f_{stn}$ (GHz)                      &      2.895        &    3.290         &    2.895         \\
               & ~~~~~~~~~~~~~~~~~~~~~~~~~~$D_{n}$ (GHz s$^{-1}$) &  -14.5$\pm$0.6    &  -13.1$\pm$0.8   &  -10.5$\pm$1.2   \\
               &         ~~~~~~~~~~~~~~~~~~$\bar{D}_{n}$ (s$^{-1}$)& -5.3$\pm$0.22    &  -4.4$\pm$0.26   &  -3.8$\pm$0.44   \\
               & RS III: $f_{str}$ (GHz)                          &      2.945        &    3.56          &    2.905         \\
               &        ~~~~~~~~~~~~~~~~~~~$D_{r}$ (GHz s$^{-1}$) &   8.6$\pm$0.8     &  5.6$\pm$0.6     &   9.7$\pm$0.8    \\
               &         ~~~~~~~~~~$\bar{D}_{r}$ (s$^{-1}$)       &   2.6$\pm$0.24    &  1.5$\pm$0.16    &   3.2$\pm$0.26   \\\hline
               &                                                  &                   &                  &                  \\
  Diagnostics  & $n_{x}$ ($10^{17}m^{-3}$)                        &      1.05         &     1.44         &    1.03          \\
               & $T$ ($10^{7}$ K)~~~~~                            &      1.85         &     1.45         &    1.38          \\
               & $n_{n}$ ($10^{17}m^{-3}$)                        &      1.035        &     1.336        &    1.035         \\
               & $B_{n}$ (G)~~~~~~~~~~~~                          &   87.4$\pm$27.2   &  80.2$\pm$26.5   &   70.4$\pm$26.4  \\
               & $\beta_{pn}$~~~~~~~~                             &       0.87        &      1.05        &    1.00          \\
               & $H_{nn}$ (km)~~~~~~~~                             & 4.3$\times10^{3}$ & 5.3$\times10^{3}$& 5.0$\times10^{3}$ \\
               & $v_{n}$ (c)~~~~~~~~~~~                           &      0.47         &     0.45         &    0.39         \\
               & $E_{n}$ (keV)~~~~~~~                             &      64.1         &     58.0         &    42.3         \\
               & $n_{r}$ ($10^{17}m^{-3}$)                        &      1.071        &     1.565        &    1.042         \\
               & $B_{r}$ (G)~~~~~~~~~~~~                          &   67.6$\pm$21.1   &  53.6$\pm$17.0   &   67.8$\pm$25.3  \\
               & $\beta_{pr}$~~~~~~~~                              &       1.50       &      2.74        &    1.09          \\
               & $H_{nr}$ (km)~~~~~~~~                             & 7.5$\times10^{3}$ & 1.4$\times10^{4}$& 5.5$\times10^{3}$ \\
               & $v_{r}$ (c)~~~~~~~~~~~                           &      0.47         &     0.46         &    0.39         \\
               & $E_{r}$ (keV)~~~~~~~                             &       64.1         &     61.0        &    42.3         \\
               & $L_{c}$ (km)~~~~~~~~~                            &       101         &     733         &    18           \\
  \noalign{\smallskip}\hline
\end{tabular}
\end{center}
\tablecomments{0.86\textwidth}{$D$ is frequency drift rate,
$\bar{D}$ is the relative frequency drift rate, $f_{0}$ is the
central frequency of the type III burst. $f_{x}$ is the separate
frequency between the normal type III bursts and the RS type III
bursts. $B$ is the magnetic field, $v$ and $E$ are the velocity
and energy of electron beams, respectively. $\beta_{p}$ is the
plasma beta value. The subscript $n$ and $r$ indicate the normal
and RS type III branches, respectively. $L_{c}$: Length of
acceleration region.}
\end{table}

The flare starts at 02:14 UT, peaks at 02:40 UT. The microwave
type III burst pairs occurred at 02:45 UT, 03:29 UT and 03:45 UT,
respectively, from several minutes to more one hour after the
flare maximum. All type III pairs are strongly right-handed
circular polarization overlapping on a long-duration broadband
microwave type IV continuum in the postflare phase. According to
the Fig.13 of Tan et al. (2010), the distance between the two
footpoints of the flaring loop is about $d\sim 2.0\times 10^{7}$
m. Assuming $t_{n}^{-1}\sim0$ and the flaring loop to be a
semicircle, $R_{c}=\frac{1}{2}d\approx1.0\times 10^{7}$ m. Then,
we the diagnostic physical parameters of the acceleration region
can be obtained which listed in Table 1. Here, we found the
estimated magnetic fields near the acceleration region are from
53.6$\pm$17.0 G to 87.4$\pm$27.2 G. These values are a little less
than the result estimated by microwave Zebra pattern structures in
the same flare (90 - 200 G, Yan et al. 2007). As we know that the
Zebra pattern emission are produced from the interior of the flare
loop while the flare primary energy-release sites are above the
flare loop. The energy of the electron beam associated to the RS
type III branch is in the range of 42 - 64 keV, which is similar
to that of the normal type III branches. This fact implies that
accelerations are possibly selfsame for the upgoing and downgoing
electron beams.

Table 1 also shows that the frequency drift rate of the normal
type III branches is higher than that of RS branches. This fact
can be explained as following: from Equation (16) we have:

\begin{equation}
\bar{D}\approx\frac{1}{2\mu_{0}k_{B}}\frac{B^{2}}{n_{st}T}\frac{v_{b}}{R_{c}}.
\end{equation}
As we know electron beam of the normal type III branches
propagates upward in a relatively rarefied plasma ($n_{stn}$)
while the electron beam of the RS branches propagates downward in
a relatively denser plasma ($n_{str}$): $n_{stn}<n_{str}$.
Equation (19) gives: $\bar{D}_{n}>\bar{D}_{r}$, the normal
branches drift faster than the RS branches.

Equation (19) shows that not only the plasma density ($n_{e}$)
affects the frequency drift rate, but also the magnetic field
strength ($B^{2}$), plasma temperature ($T$), magnetic field
configuration ($R_{c}$), and the velocity of electron beams
($v_{b}$) dominate the value of drift rates. After determining
these parameters, we can really explain the other observed
properties of microwave type III burst pairs.

Additionally, we also calculated the plasma beta value
($\beta_{p}$). The result is also listed in Table 1. Here, we find
that the plasma beta is much greater than 1 which is just
indicating that the flare primary energy-release region is a
highly-dynamic area where the magnetized plasma is unstable which
may generate plasma instability, trigger magnetic reconnection,
accelerate particles, heats the ambient plasmas, release the
magnetic energy, and produce the flaring eruptions.

According to PE mechanism, the frequency gap may imply a density
difference between start sites of normal and RS type III bursts.
With density scale length $H_{n}$, a spatial length can be
derived:

\begin{equation}
L_{c}\approx H_{nn}\cdot\frac{\delta
f}{2f_{stn}}+H_{nr}\cdot\frac{\delta f}{2f_{str}}.
\end{equation}

$L_{c}$ can be regarded as estimation of the length of
acceleration regions. The electrons are accelerated in this region
and get a relatively high energy, then trigger microwave type III
bursts outside this region. The estimated length is about 18 - 733
km. It is possible that these results are the upper length limit
of the acceleration region.

In the above estimation, we adopt the same value of $R_{c}$ for
both of normal and RS type III branches. Actually, they are
different from each other. However, Equation (2) indicates
$B\propto R_{c}^{1/2}$, and Equation (4) indicates $v_{b}\propto
R_{c}B^{-2}$, these facts show that the beam velocity ($v_{b}$) is
independent to the magnetic field strength. So, the uncertainties
of magnetic field will not affect the estimation of the velocity
and energy of the electron beams.

\section{Conclusions and Discussions}

In summary, when consider the effect of magnetic field on the
plasma density and its distribution around the flare
energy-release site, a relationship between the physical
conditions and the observing parameters of microwave type III
bursts can be established. With this relationship, we can diagnose
the flare energy-release sites directly. The diagnostic procedure
is as following,

(1) Determine the temperature ($T$) at the time of microwave type
III burst from the observation of GOES SXR at wavelength of 0.5 -
4 \AA~ and 1 - 8 \AA.

(2) Determine the plasma density ($n_{x}$) using Equation (6) from
the observing start frequency ($f_{st}$) of microwave type III
burst or the separate frequency ($f_{x}$) of microwave type III
burst pairs.

(3) Estimate the magnetic field ($B_{L}$ and $B_{H}$) by using
Equation (17) and (18) from the observing relative frequency drift
rate ($\bar{D}$), burst lifetime ($\tau$) and the curvature radius
of magnetic field lines ($R_{c}$).

(4) Estimate the velocity and energy of the energetic electron
beams using Equation (15) or (16) from the observing relative
frequency drift rate ($\bar{D}$), the curvature radius of magnetic
field lines ($R_{c}$) and the above derived magnetic field
strength ($B$).

(5) As for the microwave type III pairs, we may obtain an upper
limit of the length of acceleration region by using Equation (20)
from the observed frequency gap ($\delta f$) between the normal
and RS type III branches.

In the above diagnostic procedure, there is a key parameter
$R_{c}$. In the example of Section 4, we obtained $R_{c}$ by
simply using the EUV imaging observation. Generally, such
estimation is a bit more exact than to derive the plasma density
scale length directly from the existing plasma density models.
However, the above method is still a bit cursory. The more exact
method to derive $R_{c}$ depends on the radio spectral imaging
observations at the corresponding frequencies, such as the Chinese
Spectral Radioheliograph (CSRH, now renamed as MUSER, Yan et al.
2009). By using such new generation telescopes, we may directly
obtain the location and geometry of the source region, and get the
more exact value of the magnetic field scale length.

From the above diagnostics, we may provide basic information for
the study of flare triggering mechanism and particle acceleration.
However, the above method cannot be extended to meter or longer
wavelength type III bursts for their propagating along the open
weak magnetic fields in the higher corona. Here, the magnetic
field ($B$) becomes very weak and the curvature radius of magnetic
field lines ($R_{c}$) becomes very large,
$\frac{B^{2}}{n_{x}\mu_{0}R_{c}}\ll m_{0}g$, Equation (13)
degenerates to be the barometric format
$H_{n}\approx-\frac{k_{B}T}{m_{0}g}$. Then the above diagnostic
method will deviate from its correctness.

\begin{acknowledgements}
The authors thank Profs Stepanov A.V., Melnikov V., and Hudson H.
for their helpful suggestions and valuable discussions on this
work. B.T. acknowledges support by NSFC Grants 11273030, 11221063,
11373039, 11433006, MOST Grant 2014FY120300, CAS XDB09000000, the
National Major Scientific Equipment R\&D Project ZDYZ2009-3. H.M.
and M.K. acknowledges support by the Grant P209/12/00103 (GA CR)
and the research project RVO: 67985815 of the Astronomical
Institute AS. This work is also supported by the Marie Curie
PIRSES-GA-295272-RADIOSUN project.
\end{acknowledgements}

\label{lastpage}

\end{document}